\documentclass[article,twoside,web]{ieeecolor} 
\usepackage{tmi}
\usepackage{url}
\usepackage{amsmath,amssymb,amsfonts}
\usepackage{algorithmic}
\usepackage{graphicx}
\usepackage{textcomp}
\usepackage{booktabs}
\usepackage{multicol}
\usepackage{booktabs}
\usepackage{lipsum}
\usepackage{makecell}
\usepackage{hyperref}
\usepackage{fancyhdr}
\begin{document}
\title{\color{black} A Flexible 2.5D Medical Image Segmentation Approach with In-Slice and Cross-Slice Attention}
\author{Amarjeet Kumar, Hongxu Jiang, Muhammad Imran, Cyndi Valdes, Gabriela Leon, Dahyun Kang, Parvathi Nataraj, Yuyin Zhou, Michael D. Weiss, and Wei Shao
\thanks{Amarjeet Kumar is with the Department of Computer and Information Science and Engineering at the University of Florida, FL, USA (e-mail: amarjeetkumar@ufl.edu).}
\thanks{Hongxu Jiang is with the Department of Electrical and Computer Engineering at the University of Florida, Gainesville, FL, USA (e-mail: hongxu.jiang@ufl.edu).}
\thanks{Muhammad Imran is with the Department of Medicine at the University of Florida, Gainesville, FL, USA (e-mail: muhammad.imran@ufl.edu).}
\thanks{Cyndi Valdes is with the Department of Pediatrics at the University of Florida, Gainesville, FL, USA (e-mail: cndvaldes@gmail.com).}
\thanks{Gabriela Leon is with the College of Medicine at the University of Florida, Gainesville, FL, USA (e-mail: g.leon@ufl.edu).}
\thanks{ Dahyun Kang is with the College of Medicine at the University of Florida, Gainesville, FL, USA (e-mail: dahyun.kang@ufl.edu).}
\thanks{Parvathi Nataraj is with the Department of Pediatrics at the University of Florida, Gainesville, FL, USA (e-mail: parvathinataraj@yahoo.com).}
\thanks{Yuyin Zhou is with the Department of Computer Science and Engineering, University of California, Santa Cruz,  CA, USA (e-mail: yzhou284@ucsc.edu).}
\thanks{Michael D. Weiss is with the Department of Pediatrics at the University of Florida, Gainesville, FL, USA (e-mail: weissmd@peds.ufl.edu).}
\thanks{Wei Shao is with the Department of Medicine, Department of Electrical and Computer Engineering, and the Intelligent Clinical Care Center, University of Florida, Gainesville, FL, USA (e-mail: weishao@ufl.edu).}
\thanks{Yuyin Zhou, Michael D. Weiss, and Wei Shao contributed equally as senior authors.}
} \maketitle
\thispagestyle{fancy}

\begin{abstract}
Deep learning has become the de facto method for medical image segmentation, with 3D segmentation models excelling in capturing complex 3D structures and 2D models offering high computational efficiency. However, segmenting 2.5D images, which have high in-plane but low through-plane resolution, is a relatively unexplored challenge. While applying 2D models to individual slices of a 2.5D image is feasible, it fails to capture the spatial relationships between slices. On the other hand, 3D models face challenges such as resolution inconsistencies in 2.5D images, along with computational complexity and susceptibility to overfitting when trained with limited data. In this context, 2.5D models, which capture inter-slice correlations using only 2D neural networks, emerge as a promising solution due to their reduced computational demand and simplicity in implementation.
In this paper, we introduce CSA-Net, a flexible 2.5D segmentation model capable of processing 2.5D images with an arbitrary number of slices through an innovative Cross-Slice Attention (CSA) module. This module uses the cross-slice attention mechanism to effectively capture 3D spatial information by learning long-range dependencies between the center slice (for segmentation) and its neighboring slices. Moreover, CSA-Net utilizes the self-attention mechanism to understand correlations among pixels within the center slice. We evaluated CSA-Net on three 2.5D segmentation tasks: (1) multi-class brain MRI segmentation, (2) binary prostate MRI segmentation, and (3) multi-class prostate MRI segmentation. CSA-Net outperformed leading 2D and 2.5D segmentation methods across all three tasks, demonstrating its efficacy and superiority. Our code is publicly available at \url{https://github.com/mirthAI/CSA-Net}. 
\end{abstract}


\section{Introduction}
\label{sec:introduction}
\IEEEPARstart{M}{edical} image segmentation is essential in computer-assisted diagnosis, treatment planning, surgical navigation, and image-guided interventions~\cite{siddique2021u}. 
2D image segmentation involves delineating regions of interest within 2D image slices, such as blood vessel segmentation in retinal images~\cite{gegundez2021new}, cell segmentation in microscopy images~\cite{al2018deep}, and lung segmentation in chest X-rays~\cite{kim2021automatic}. In contrast, 3D image segmentation involves segmenting regions of interest in 3D from volumetric images, such as abdominal organ segmentation in CT~\cite{gibson2018automatic} and brain tumor segmentation in MRI~\cite{havaei2017brain}.
Since the introduction of the U-Net model~\cite{al2018deep}, numerous deep learning-based segmentation methods have been developed for a wide range of 2D and 3D medical image segmentation tasks. 

\begin{figure}[!hbt]
\centerline{\includegraphics[width=\columnwidth]{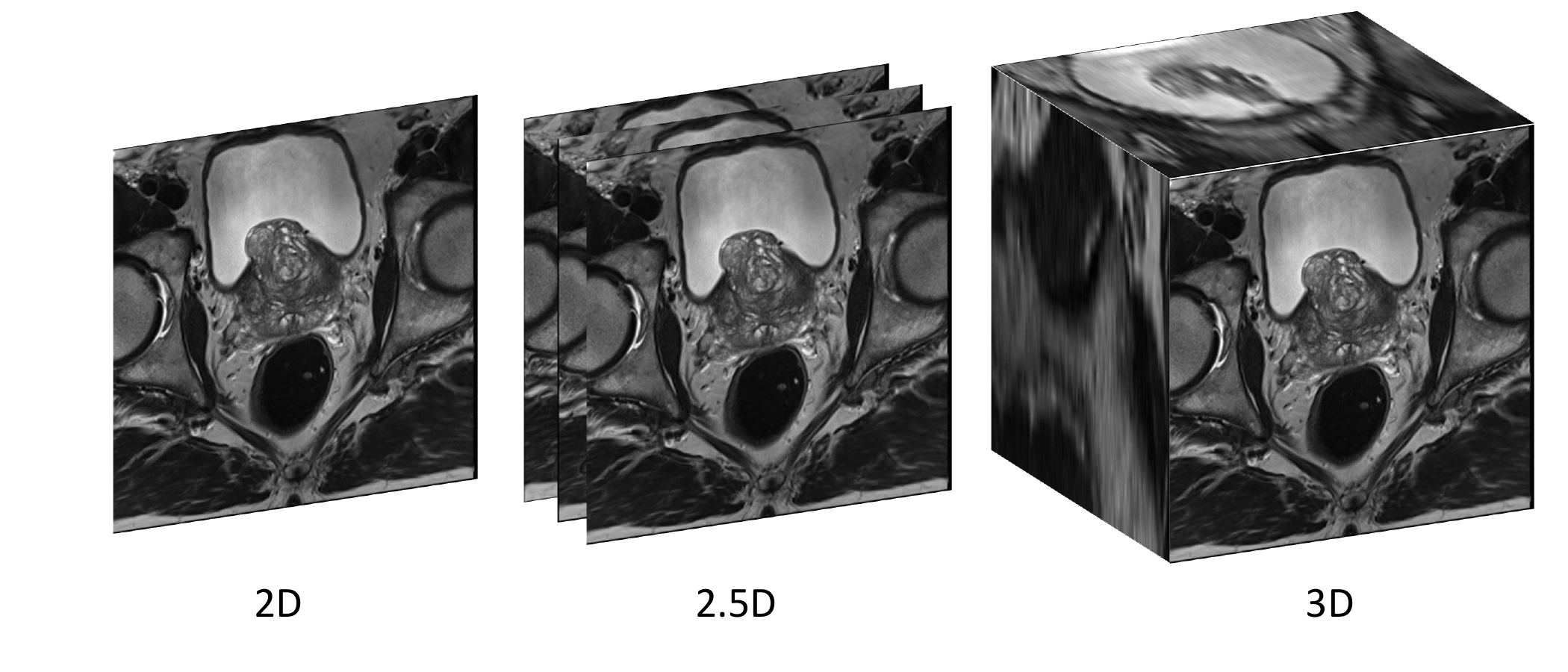}}
\caption{Illustration of the inputs to 2D, 2.5D, and 3D segmentation models. }
\label{fig:2.5D_vs3D}
\end{figure}

An important, yet less explored, segmentation challenge is the 2.5D image segmentation problem. This task aims to segment objects within a 2.5D image, which consists of a sequence of  2D image slices with high in-plane resolution but low through-plane resolution (see Fig.~\ref{fig:2.5D_vs3D}).
In other words, for 2.5D images, spatial resolutions across the three dimensions are not uniform, with the $z$-axis exhibiting a lower resolution compared to the $x$ and $y$ axes.
The segmentation of 2.5D images poses challenges due to their inter-slice discontinuities and partial volume effects which are caused by limited resolution in the $z$-axis. 2D segmentation models are commonly used to segment 2.5D images. This is done by segmenting each individual 2D slice within the 2.5D image using a 2D segmentation network. Although, this approach is simple and computationally efficient, it fails to fully leverage the spatial context across slices, leading to potential segmentation errors in regions where objects span multiple slices.
On the other hand, while 3D segmentation models can integrate contextual information across slices in 3D, they have several limitations when applied to 2.5D segmentation problems. First, the increased complexity of 3D models requires significantly higher computational resources compared to 2D models. Second, given that many medical datasets are relatively small, the increased number of parameters in 3D models raises the risk of overfitting, potentially compromising the model's generalization ability. Third, applying 3D methods to 2.5D images is not straightforward. This is because the number of slices in a 2.5D image can vary and be fewer than 32, which is the minimum required by most 3D segmentation methods.  These discrepancies make it challenging to design a 3D segmentation model that fits 2.5D image volumes.

Recently, 2.5D segmentation methods have emerged as a promising solution to bridge the gap between 2D and 3D segmentation techniques~\cite{xia2018bridging,hu20182,zhang2019multiple}. These methods utilize 2D neural networks to ensure computational efficiency and ease of adaptation to images with varying through-plane resolution, while also learning inter-slice relationships for a more holistic understanding of 3D image features.
A common approach, used in 2.5 segmentation models to capture nearby slice information, involves the stacking of multiple 2D slices into a single 2D image with multiple channels~\cite{han2021liver,zhang2019multiple,soerensen2021deep}.
However, simply concatenating adjacent slices as model input, typically ranging from 3 to 7 slices~\cite{soerensen2021deep}, limits effective integration of inter-slice spatial context because it treats neighboring slice information superficially as depth information. 
Alternatively, cross-slice attention mechanisms are used~\cite{hung2022cat,hung2024csam} to learn correlations between all slices in a 2.5D volume.
For instance, CAT-Net~\cite{hung2022cat} processes all slices simultaneously to learn slice-level attention scores, which require the input 2.5D volume to have a fixed number of slices. This therefore means that CAT-Net is not applicable to 2.5D volumes with a varying number of 2D slices. Additionally, CAT-Net learns correlations only at the slice level, and therefore is unable to capture correlations between image regions within individual slices and across different slices.

In this paper, we introduce CSA-Net, a 2.5D segmentation model that incorporates a novel Cross-Slice Attention (CSA) module. CSA-Net takes a center slice and its two neighboring slices as input to predict the segmentation map for the center slice. Specifically, the CSA module employs the cross-attention mechanism~\cite{wang2018non} to capture the correlation between regions in the center slice and regions in its neighboring slices. Additionally, CSA-Net utilizes the self-attention mechanism to learn correlations between different regions within the center slice. By combining outputs from different attention blocks, CSA-Net generates a unified feature map encapsulating crucial information from all three slices. This significantly enhances the segmentation accuracy on the center slice.
We compared CSA-Net with leading 2D and 2.5D segmentation methods on three different 2.5D image segmentation tasks: multi-class brain MRI segmentation in our private dataset, binary prostate MRI segmentation in a public dataset, and multi-class prostate MRI segmentation in another public dataset. CSA-Net consistently outperformed existing methods across all three datasets, demonstrating its superiority in leveraging both in-slice and cross-slice spatial relationships to achieve more accurate and reliable segmentation results. 
In summary, this paper makes the following key contributions:
\begin{itemize}
\item We introduced CSA-Net, a 2.5D segmentation model incorporating a cross-slice attention module and an in-slice attention module, which effectively captures both in-slice and cross-slice spatial relationships.
\item We demonstrated CSA-Net's superior performance over existing 2D and 2.5D segmentation methods across three different segmentation tasks.
\item We validated the advantages of 2.5D methods over 2D methods in segmenting 2.5D images, promoting the exploration of more advanced 2.5D segmentation methods.
\end{itemize}

\section{Related Work}
\subsection{Transformer Based Image Segmentation Models}

The success of transformer models in the field of natural language processing has prompted research into their applicability across various domains, notably in computer vision. A key development for the research was the introduction of the Vision Transformer (ViT)~\cite{dosovitskiy2020image}, which marked the first transformer-based model to outperform convolutional networks in image classification. ViT innovatively employs a patch-based attention mechanism, in which an image is divided into fixed-size patches that are each treated as an individual token. This allows the model to dynamically focus on and relate different parts of an image through self-attention, significantly enhancing its ability to capture complex visual relationships in a computationally efficient manner.

Building on the foundational success of ViT in image classification, researchers have extensively explored the integration of ViT with the U-Net architecture for image segmentation tasks. This integration capitalizes on ViT's global contextual understanding and U-Net's precise localization capabilities. TransUnet~\cite{chen2021transunet} stands out as the pioneering image segmentation architecture for leveraging vision transformers. Subsequent advancements include nnFormer~\cite{zhou2023nnformer}, which enhances the methodology by interleaving convolution with self-attention. Recently, the Swin Transformer~\cite{liu2021swin}, a hierarchical vision transformer using shifted windows to efficiently capture both local and global visual information, has inspired the development of image segmentation models like SwinUNet~\cite{cao2022swin} and Swin UETR~\cite{hatamizadeh2021swin}. Transformer-based segmentation models have been successfully applied to various medical imaging tasks such as brain volume segmentation~\cite{zhang2022tw,chen2023transattunet,yang20243d} and prostate segmentation~\cite{hung2022cat,ding2023multi}.
 
\subsection{Cross-Attention Mechanism}
The cross-attention mechanism, initially used in the Transformer decoder for machine translation~\cite{vaswani2017attention}, enables each token in one sequence to selectively attend to all tokens in another sequence. Mathematically, computation of the attention scores involves the query coming from one sequence, and the key-value pairs coming from another sequence. This mechanism is particularly useful for tasks that involve multiple input modalities or require alignment between different sequences. Cross-attention has been successfully applied to applications such as pairwise image registration~\cite{song2021cross}, where the inputs were two different images. Additionally, it has been integrated into fusion networks for segmentation tasks~\cite{zheng2023casf}, where information from multiple sources and modalities were integrated for more accurate segmentation. Recently, the cross-attention mechanism was integrated within the skip connections of a 3D segmentation network to enhance the decoder's performance~\cite{zhou2021cross}. These advancements demonstrate the versatility and effectiveness of the cross-attention mechanism across various medical imaging applications.

\subsection{2.5D Segmentation Methods}
 
Several 2.5D image segmentation methods have been developed to capture inter-slice relationships at a lower computational cost than 3D methods. A feasible approach, adopted by several models~\cite{li2021acenet,yu2018recurrent,wang2019benchmark,duan2019automatic}, involves concatenating multiple (usually 3, 5, or 7) consecutive image slices into a 2D image with multiple channels. This combined image is then used by a 2D segmentation network to segment regions of interest in the middle slice. However, directly concatenating neighboring slices as a multi-channel 2D image complicates information extraction for individual slices, causing misalignment between consecutive slices. This could possibly introduce inaccuracies into the segmentation model.
To mitigate these issues, Recurrent Neural Networks (RNNs) have been utilized, which treat 2.5D images as temporal sequences~\cite{yu2018recurrent, yang2018towards}. Nevertheless, due to their sequential nature, the computational cost for training remains considerably high.

To further reduce the computational cost, a shift towards the adoption of attention mechanisms in 2.5D segmentation networks has been made to extract features related to inter-slice relationships. Such mechanisms allow the models to selectively focus on relevant regions within neighboring slices, thereby enhancing the accuracy and robustness of segmentation for the center slice. For instance, the CAT-Net and CSAM models~\cite{hung2022cat,hung2024csam} employ a slice-level cross-slice attention module to learn a correlation coefficient between each pair of 2D slices. However, this method focuses solely on inter-slice correlation at the slice level and fails to capture correlations between different image regions. Moreover, both the CAT-Net and CSAM models require a fixed number of slices for input 2.5D images, making them unsuitable for images with varying through-plane resolutions, which is a common scenario. Additionally, since all slices are processed simultaneously, this approach constrains the models' scalability due to the substantial memory requirements.

\section{Methods}

\begin{figure*}[!hbt]
\centerline{\includegraphics[width=\linewidth]{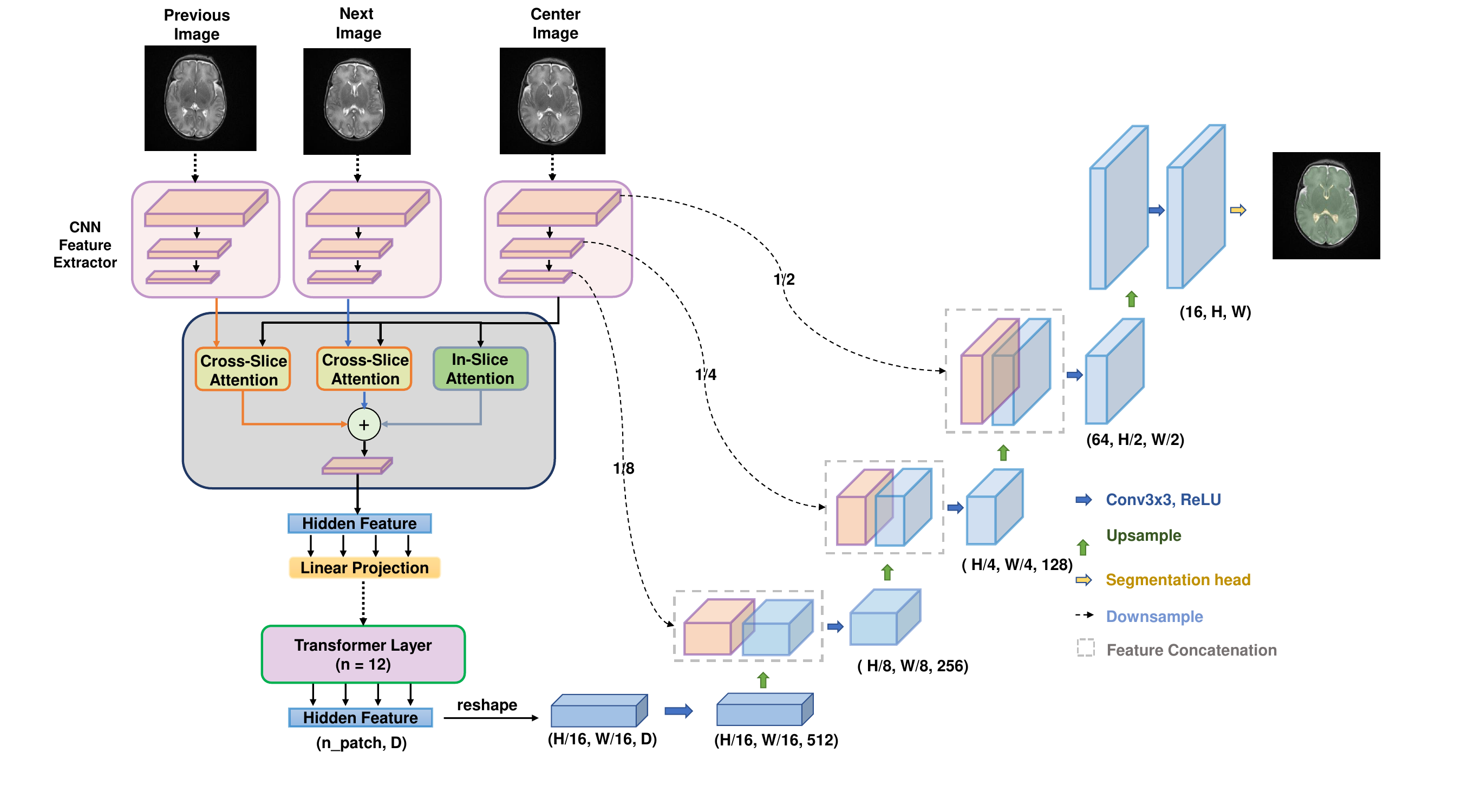}}
\vspace{-2em}
\caption{Overview of CSA-Net's architecture. 
}
\label{fig:overview}
\end{figure*}

This paper introduces CSA-Net, a novel medical image segmentation model designed for 2.5D images. Figure~\ref{fig:overview} presents an overview of the architecture of CSA-Net. The model takes three consecutive 2D slices as input: a center slice and its two neighboring slices (previous and next), producing the segmentation of the center slice. We designed two key modules to enhance the segmentation results. First, the cross-slice attention module (detailed in Section~\ref{sec:csa}) enables the center slice to utilize information from its neighboring slices via pixel-level cross-attention. Second, the in-slice attention module (detailed in Section~\ref{sec:in-slice-att}) learns correlations between different regions within the center slice via pixel-level self-attention. Recognizing the complementary nature of the cross-slice and in-slice attention mechanisms, we integrated the aggregated outputs from these attention modules into a vision transformer encoder, thereby further enhancing representation learning. Subsequently, the decoder, which is comprised of a series of transposed convolutional layers, is employed to produce the final segmentation map.

\subsection{Cross-Slice Attention}   
\label{sec:csa}
\begin{figure}[!hbt]%
\centerline{\includegraphics[width=\columnwidth]{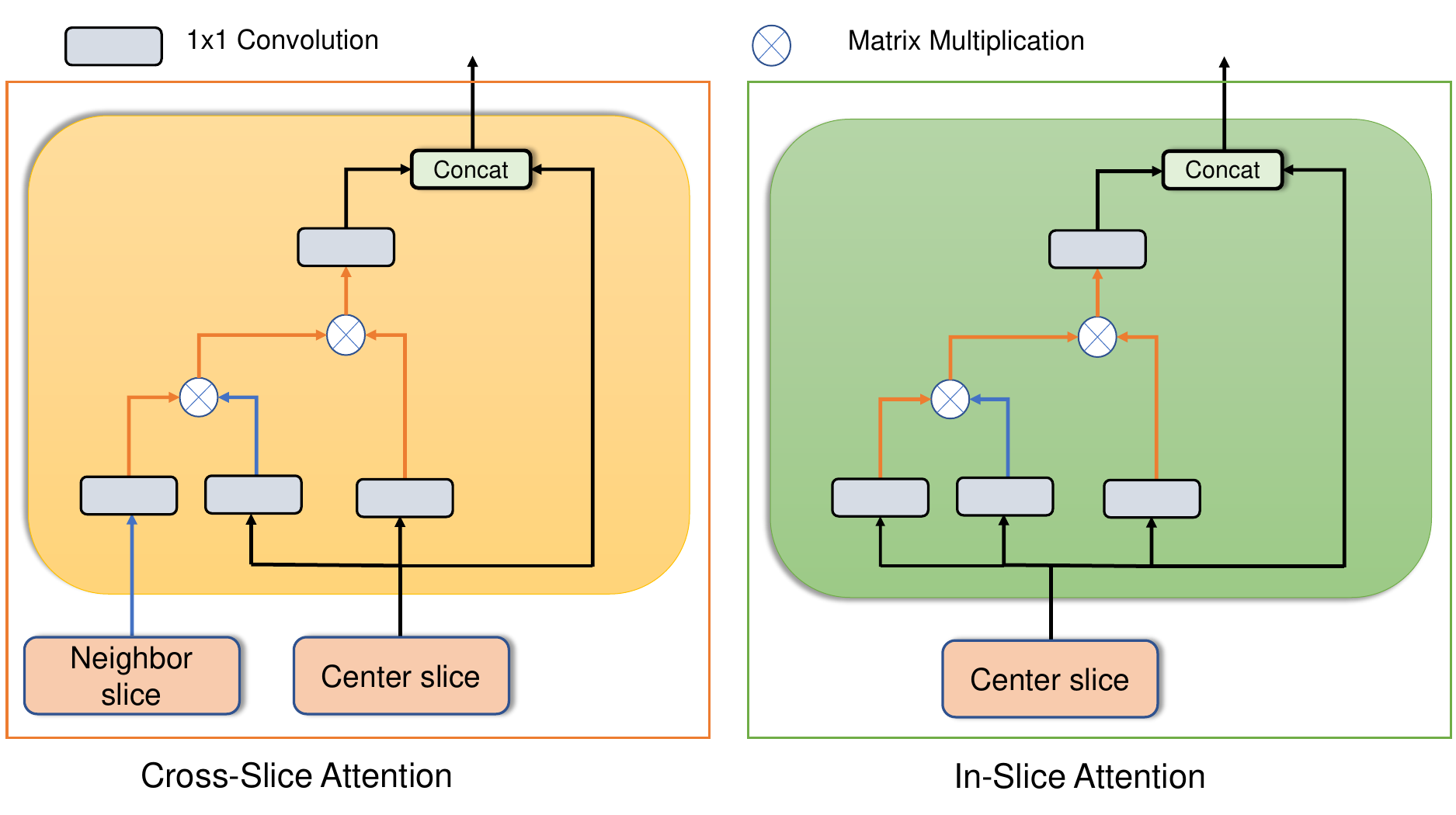}}
\caption{Overview of our Cross-Slice attention (left) and In-Slice attention (right) architecture. }
\label{fig:cross_slice_attention}
\end{figure}

We developed a Cross-Slice Attention (CSA) module (see Fig.~\ref{fig:cross_slice_attention} to learn the correlation between different image regions across three consecutive slices: the previous, the center, and the next slices, each of size $H\times W$. The input to the CSA module consists of feature maps $f_c$ and $f_n$ extracted from the center slice and one of its neighboring slices using convolutional neural networks, as detailed in Section~\ref{sec:feature_extract}. The size of the feature map for each slice is $h\times w\times C$, where the downsampling ratio during feature extraction is defined as $\frac{H}{h} = \frac{W}{w} = 2^{r}$, with $r > 1$, and $C$ denotes the number of channels in each feature map.
Two $1\times 1$ convolutional layers, parameterized by $\phi$ and $\psi$ (with weight matrices $W_{\phi}$ and $W_{\psi}$), were applied to the feature map of the center slice, resulting in two image feature maps of size $h\times w \times \frac{C}{2}$, serving as the key-value pair in the CSA module. 
A $1\times 1$ convolutional layer parameterized by $\theta$ with weight matrix $W_\theta$ was applied to the feature map of the neighboring slice, resulting in a feature map of size $h\times w \times \frac{C}{2}$ to serve as the query.

We compute an attention score between each pixel in the output feature map of the center slice and each pixel in the output feature map of the neighboring slice. The output from the attention module is of size $h\times w \times \frac{C}{2}$. To ensure the input and output of the CSA module have the same dimension, a $1\times 1$ convolutional layer with parameter matrix $W_g$ was applied to double the number of channels. To summarize, the output from the CSA module is given by
\begin{equation}
CSA(f_c,f_n) = \left ( \text{softmax}(f_{n}^{T}W^T_\theta f_{c}W_{\phi}) f_{c}W_{\psi} \right) W_g,
\end{equation}
where $W_\theta, W_\phi, W_\psi$ are of dimension $C \times \frac{C}{2}$ and $W_g$ is of dimension $\frac{C}{2}\times C$. During the computation, all feature maps were flattened into a spatial size of $hw$, and reshaped to a spatial size $h\times w$ after the computation.

To learn different types of relationships between the center slice and the neighboring slice, we implemented the cross-slice attention module using multiple attention heads, similar to the Transformer network~\cite{vaswani2017attention}.
The outputs from all cross attention heads are concatenated into a single image feature map of size $h\times w \times C$, which is then concatenated with the feature map of the center slice to yield the final output.

\subsection{In-Slice Attention}
\label{sec:in-slice-att}
In addition to capturing the cross-slice correlation, we implemented an In-Slice Attention (ISA) module (see Fig.~\ref{fig:cross_slice_attention}) which employs the self-attention mechanism to learn the correlation between different regions within the center slice. The ISA module receives an input from the feature map $f_c$, extracted from the center slice via the convolutional feature extractor described in Section~\ref{sec:feature_extract}.
Three $1\times 1$ convolutional layers, denoted as $\alpha(\cdot)$, $\beta(\cdot)$, and $\gamma(\cdot)$ with weight matrices $W_\alpha, W_\beta,$ and $W_\gamma$ respectively, are applied to $f_c$, producing the query, key, and value matrices essential for computing the attention scores. This is subsequently used to calculate the attention output. Mathematically, the output of the in-slice attention module can be formulated as follows:
\begin{equation}
ISA(f_c) = \left( \text{softmax} (f^T_c W_\alpha^{T} f_c W_\beta) f_cW_\gamma \right) W_\epsilon
\end{equation}
where $W_\epsilon$ represents the weight matrix of a $1 \times 1$ convolutional layer designed to transform the output dimensionality of the attention module from $h \times w \times \frac{C}{2}$ back to $h \times w \times C$, ensuring consistency in feature map dimensions.
Similarly, multiple attention heads are utilized to capture different types of pixel correlations.

\subsection{Architecture of CSA-Net}
\label{sec:architecture}

\subsubsection{\textcolor{black}Feature Extractor} 
\label{sec:feature_extract}
The three input image slices (the previous slice, the center slice, and the next slice), each of dimension $H \times W$, for CSA-Net, are initially passed through a CNN encoder (i.e., the ResNet-50 model) for feature extraction. This process yields image feature maps of dimensions $h \times w \times C$, and we set $\frac{H}{h} = \frac{W}{w} = 16$ and $C = 1024$ in this paper. Subsequently, the feature maps of the three slices are input into the cross-slice and in-slice attention modules (Sections~\ref{sec:csa} and~\ref{sec:in-slice-att}).

\subsubsection{\textcolor{black}{Attention Aggregation for 2.5D Image Understanding}} 
We employed two CSA modules to capture 2.5D spatial information: one to capture the correlation between the center slice and the previous slice, and the other to capture the correlation between the center slice and the next slice. Additionally, we used an ISA module to capture the long-range dependencies within the center slice. The outputs from these three attention modules are concatenated along the channel dimension to generate the final attention feature output. This combined feature attention map is then passed through a $1 \times 1$ convolutional layer to reduce the number of channels to $C$, which is the same as the number of channels in the feature map $f_c$.

\subsubsection{\textcolor{black}{Vision Transformer Encoder}}
We utilized a pre-trained 12-layer Vision Transformer (ViT) to augment the extraction of global information from the preceding cross-slice and in-slice attention modules. In contrast to the original ViT, which processes $16 \times 16$ image patches, we chose a patch size of $1 \times 1$, given that our input to ViT is already a low-dimensional feature map. Subsequently, the ViT's output is reshaped to match its input dimensions, namely, $h \times w \times C$. Following this, the encoded feature map is then fed into the decoder to generate the segmentation map.

\subsubsection{Decoder} The decoder entirely relies on convolutional neural networks, comprising four decoder blocks. Each block employs both transposed convolution and regular convolutional layers, in addition to residual connections. This setup upsamples the feature map by a factor of 2 at each level, progressively refining the spatial resolution. The concluding convolutional layer, equipped with a filter size of $1\times1$, generates the final segmentation map with dimensions $H\times W$.

\subsubsection{\textcolor{black}{Loss Function}}
 We used a combination of the cross-entropy and the Dice loss for the training:

\begin{equation}
\mathcal{L}  = 0.5 \times \mathcal{L}_{CE} + 0.5 \times \mathcal{L}_{DSC},
\end{equation}
where $\mathcal{L}_{CE}$ denotes the cross entropy loss and $\mathcal{L}_{DSC}$ denotes the dice loss.
\section{Experimental Design}
\subsection{Dataset}

\subsubsection{\textcolor{black}{Brain Dataset}} 
This study, which has been approved by the University of Florida's Institutional Review Board, conducted a retrospective chart review of 57 T2-weighted brain MRI image volumes of infants born with gestational ages ranging from 22 to 29 weeks. Brain MRIs were acquired using a 1.5 Tesla Magnetom MRI (Avanto and Aera, Siemens Medical Solutions USA, Inc, Malvern, PA, USA) or a 3.0 Tesla MAGNETOM MRI (Vario, Prisma, and Skyra, Siemens). Each MRI contained 25-30 2D axial slices of dimensions $(320 \times 320)$ with an isotropic in-plane resolution of 0.50 mm and a through-plane resolution of 10 mm. We manually segmented the brain and the ventricles from each MRI in the ITK-Snap software. We randomly selected 42 MR images for training and 15 MR images for testing.

\subsubsection{\textcolor{black}{Promise12 Dataset}} 
Our study included 80 T2-weighted MR images from the publicly accessible Promise12 Dataset~\cite{litjens2014evaluation}, each accompanied by manual binary segmentation of the prostate capsule. These MRI images were obtained using either a 1.5 Tesla Magnetom MRI scanner (Siemens) or a 3.0 Tesla Magnetom MRI scanner (General Electric and Siemens). Each 2D MRI slice is of size $320 \times 320$, with in-plane resolutions ranging from 0.25 mm to 0.625 mm. The through-plane resolution of the MR images ranged between 2.2 mm and 4 mm. For this dataset, we randomly selected 50 samples for the training set and 30 for the testing set.

\subsubsection{\textcolor{black}{ProstateX Dataset}} 
We included 98 T2-weighted MRI images from the publicly available ProstateX Dataset~\cite{armato2018prostatex} in this study. The MRI images were obtained using Siemens MAGNETOM Trio and Skyra 3T MR scanners with an isotropic in-plane resolution of 0.50 mm and a through-plane resolution of 3.0-4.0 mm. Each MRI image consists of 18 to 22 2D axial slices, with a standardized image size of 
$384 \times 384$. Each MRI image had an expert-annotated segmentation map for four different classes: transition zone, peripheral zone, urethra, and anterior fibromuscular stroma. We randomly selected 68 MRI images for training and 30 MRI images for testing.

\subsection{Methods for Comparison}
We compared CSA-Net with several leading segmentation methods: UNet~\cite{ronneberger2015u}, Dilated Residual UNet (DRUNet)~\cite{vesal2022domain}, SegResNet~\cite{hsu2021brain}, and TransUNet~\cite{chen2021transunet}. UNet is known for its effectiveness in biomedical image segmentation tasks, which served as a baseline for comparison due to its widespread use and simplicity. The Dilated Residual UNet architecture integrates dilated convolutions and residual connections to efficiently capture multi-scale contextual information. 
SegResNet combines Unet-based architecture and ResNet-like blocks in each encoder layer, utilizing both residual and skip connections to capture fine details and a holistic understanding of image context by integrating local and global information.
TransUNet leverages the power of vision transformer networks to capture long-range dependencies in images. These methods were originally proposed for 2D segmentation tasks. We also implemented a 2.5D version for each of them by using three consecutive slices as input to the model for segmentation of the center slice. We termed these 2.5D methods as: 2.5D UNet, 2.5D DRUNet, 2.5D SegResNet, and 2.5D TransUNet.

\subsection{Evaluation Metrics}
For the evaluation of segmentation performance, we employed two key metrics: Dice Coefficient and HD95 (95th percentile of Hausdorff Distance). The Dice Coefficient quantifies the relative overlap between the ground truth ($G$) and the predicted segmentation ($P$):
\begin{equation}
DSC(G,P) = \frac{2\times\left| G \cap P \right |}{\left|G\right|+ \left|P\right|}
\end{equation}
where $\left|G\right|$ and $\left|P\right|$ denote the number of positive pixels in the segmentation images of the ground truth and the prediction, respectively. $\left| G \cap P \right|$ represents the number of pixels that are correctly classified as positive by the prediction. The Dice coefficient ranges from 0 to 1, where a value of 0 indicates no overlap between the ground truth and the prediction, while a value of 1 indicates a perfect overlap.

Hausdorff distance (HD) measures the maximum distance between two sets of points. Specifically, HD95 represents the 95th percentile of the Hausdorff Distance, offering a robust evaluation metric. The Hausdorff distance can be defined as:
\begin{equation}
HD(G,P) = \max\left\{ \max_{g \in G} \min_{p \in P} d(g,p), \max_{p \in P} \min_{g \in G} d(p,g) \right\}
\end{equation}
where $d(g,p)$ represents the Euclidean distance between point $g$ in the ground truth set $G$ and point $p$ in the predicted segmentation set $P$.

\subsection{Implementation Details}
As a preprocessing step, all image slices were resized to a dimension of $256\times256$ pixels and normalized to the range of [0,1] using min-max normalization, with the 1st and the 99th percentiles of the pixel intensity distribution serving as the minimum and maximum values, respectively. Data augmentation techniques, including random flipping, translation, and brightness shifts, were applied across all datasets. For the Prostatex Dataset, characterized by low image contrast, we utilized the Contrast Limited Adaptive Histogram Equalization (CLAHE) method to significantly enhance image contrast.

All models were run on a high-performance computer node equipped with an NVIDIA A100 GPU, 40 GB of RAM, and 8 CPU cores. For training, the Adam optimizer was employed, set with a learning rate of $10^{-3}$ and a weight decay of $10^{-5}$. We adopted a batch size of 8 for both the Brain and Promise12 datasets, while for the ProstateX dataset, the batch size was set to 16. The models were trained for 50 epochs.

\section{Results}
\subsection{Quantitative Evaluation}
\subsubsection{\textcolor{black}{Brain Dataset}} 
Table~\ref{tb:brain_res} presents the Dice Coefficients and Hausdorff distances for various segmentation methods, averaged over 15 test cases. Our CSA-Net model surpassed all others in both Dice coefficient and Hausdorff distance, showcasing its superior segmentation accuracy and boundary precision. Notably, 2.5D models consistently outperformed their 2D counterparts, highlighting the benefits of leveraging inter-slice correlations in 2.5D segmentation tasks. Compared to the second-best model, 2.5D TransUNet, CSA-Net significantly enhanced the Dice Coefficient for the brain from 0.957 to 0.967 and markedly reduced the Hausdorff Distance from 0.92 mm to 0.68 mm by integrating cross-slice and in-slice attention modules. For the ventricles, CSA-Net improved the Dice Coefficient from 0.809 to 0.826 and lowered the Hausdorff Distance from 2.38 mm to 2.12 mm, indicating substantial improvements in segmentation accuracy.
These advancements in the segmentation of the brain and ventricles facilitate more accurate tracking of their volumes over time, offering significant implications for the analysis of brain development.

\begin{table}[!hbt]
  \centering
  \caption{Dice coefficients and Hausdorff distances (mm) in the segmentation of the brain and ventricles.}
    \label{tb:brain_res}
  \begin{tabular}{p{2.0cm} *{5}{c}} 
    \toprule
    \thead{Method} & \thead{DSC\\Brain $\uparrow$} & \thead{DSC\\Ventricles $\uparrow$} & \thead{HD95\\Brain  $\downarrow$} & \thead{HD95 \\Ventricles  $\downarrow$}  \\
    \midrule
    2D Unet & 0.937 & 0.782 & 1.142 & 2.75  \\
    2.5D Unet & 0.939 & 0.785 & 1.138 & 2.73 \\
    2D DRUnet & 0.941 & 0.791 & 1.082 & 2.66   \\
    2.5D DRUnet & 0.946 & 0.794 & 1.041 & 2.62   \\
    2D SegResNet & 0.947 & 0.794 & 1.036 & 2.59   \\
    2.5D SegResNet & 0.951 & 0.798 & 0.977 & 2.48   \\
    2D TransUnet & 0.951 & 0.800 & 0.946 & 2.43   \\
    2.5D TransUnet & 0.957 & 0.809 & 0.921 & 2.38   \\
    \midrule[0.1pt]
    CSA-Net & \textbf{0.967} & \textbf{0.826} & \textbf{0.682} & \textbf{2.12}   \\
    \bottomrule
  \end{tabular}
\end{table}

In addition, we also trained the  3D Swin UNetR~\cite{hatamizadeh2021swin} model for brain segmentation. Since our brain MRI contains fewer than 32 slices, we upsampled each MRI in the $z$ direction to double the number of slices. The resulting Dice and HD95 scores for the brain were 0.941 and 1.16 mm, respectively. For the ventricles, the Dice and HD95 scores were 0.751 and 2.93 mm, respectively. The suboptimal performance of the Swin UNetR model suggests that 3D segmentation models may not be well-suited for 2.5D segmentation tasks, possibly due to the partial volume effect introduced by upsampling. Therefore, in this study, we mainly focus on comparison with existing 2D and 2.5D methods as detailed below.

\subsubsection{\textcolor{black}{Promise12 Dataset}}
Results presented in Table~\ref{tb:pros_res} demonstrate that our CSA-Net model surpassed all competing models in binary prostate segmentation on the Promise12 dataset, achieving superior performance in terms of Dice coefficients and Hausdorff distances. Specifically, CSA-Net enhanced the Dice coefficient to 0.921 from 0.910 and reduced the Hausdorff distance to 1.06 mm from 1.14 mm, compared to the second-best performing method, 2.5D TransUNet. This underscores the effectiveness of our model. Notably, 2.5D methods consistently outperformed their 2D counterparts, and the transformer-based methods (including CSA-Net and TransUNet), which provide a more comprehensive view of the image, surpassed methods relying solely on convolutional networks (UNet, DRUNet, and SegResNet). The improved prostate segmentation is beneficial for enhancing MRI-targeted prostate biopsy and early prostate cancer detection on MRI.

\begin{table}[!hbt]
  \centering
  \caption{Dice coefficients and Hausdorff distances (mm) in the segmentation of the prostate capsule.}
  \label{tb:pros_res}
  \begin{tabular}{p{2.0cm} *{2}{c}} 
    \toprule
    \thead{Method} & \thead{DSC $\uparrow$} & \thead{HD95  $\downarrow$}   \\
    \midrule
    2D UNet & 0.872 & 1.57 \\
    2.5D UNet & 0.881 & 1.46 \\
    2D DRUnet & 0.884 & 1.30 \\
    2.5D DRUNet & 0.892 & 1.25 \\
    2D SegResNet & 0.899 & 1.22 \\
    2.5D SegResNet & 0.903 & 1.20 \\
    2D TransUNet & 0.908 & 1.19 \\
    2.5D TransUNet & 0.910 & 1.14 \\
    \midrule[0.1pt] 
    CSA-Net & \textbf{0.921} & \textbf{1.06} \\
    \bottomrule
  \end{tabular}
\end{table}

\subsubsection{\textcolor{black}{ProstateX Dataset}}

\begin{table*}[t]
  \centering
  \caption{Dice coefficients and Hausdorff distances (mm) in the segmentation of the transition Zone, peripheral zone, urethra, and anterior fibromuscular stroma of the prostate.}
  \label{prosx_res}
  \begin{tabular}{p{2.0cm} *{10}{c}} 
    \toprule
    \thead{Method} & \thead{Average\\DSC $\uparrow$} & \thead{DSC\\Transition\\Zone $\uparrow$} & \thead{DSC\\Peripheral\\Zone $\uparrow$} & \thead{DSC\\Urethra $\uparrow$} & \thead{DSC\\Anterior $\uparrow$} & \thead{Average\\HD95 $\downarrow$} & \thead{HD95\\Transition\\Zone  $\downarrow$} & \thead{HD95\\Peripheral\\Zone  $\downarrow$} & \thead{HD95\\Urethra  $\downarrow$} & \thead{HD95\\Anterior  $\downarrow$} \\
    \midrule
    2D UNet & 0.579 & 0.809 & 0.651 & 0.555 & 0.300 & 4.04 & 3.14 & 4.76 & 2.73 & 5.51 \\
    2.5D Unet & 0.601 & 0.823 & 0.676 & 0.587 & 0.316 & 3.80 & 2.93 & 4.23 & 2.69 & 5.38 \\
    2D DRUNet & 0.587 & 0.821 & 0.654 & 0.548 & 0.323 & 3.85 &2.76  &4.75  & 2.79 & 5.10 \\
    2.5D DRUNet & 0.594 & 0.829 & 0.663 & 0.558 & 0.326 & 3.70 & 2.67 & 4.37 & 2.71 & 5.08 \\
    2D SegResNet & 0.623 & 0.845 & 0.696 & 0.589 & 0.363 & 3.36 &2.39  & 3.87  & 2.41 & 4.77 \\
    2.5D SegResNet & 0.637 & 0.851 & 0.709 & 0.604 & 0.384 & 3.13 & 2.26 & 3.64 & 2.21 & 4.41 \\
    2D TransUNet & 0.631 & 0.848 & 0.704 & 0.593 & 0.377 & 3.21 & 2.29 & 3.69 & 2.25 & 4.63 \\
    2.5D TransUNet & 0.647 &\textbf{0.855} & 0.713 & 0.615 & 0.406 & 2.97 & \textbf{2.21} & 3.57 & 2.08 & 4.02 \\
    \midrule[0.1pt]
    CSA-Net & \textbf{0.659} & 0.851& \textbf{0.720} & \textbf{0.653} & \textbf{0.413} & \textbf{2.70} & 2.26 & \textbf{3.43} & \textbf{1.43} & \textbf{3.71} \\
    \bottomrule
  \end{tabular}
\end{table*}

Results in Table~\ref{prosx_res} show that our CSA-Net outperformed all other methods in terms of the Dice coefficient and Hausdorff distance metrics, averaged across the four classes: transition zone, peripheral zone, urethra, and anterior fibromuscular stroma. In comparison to the second-best performing method, 2.5D TransUNet, CSA-Net improved the average Dice coefficient from 0.647 to 0.659 and reduced the average Hausdorff distance from 2.97 mm to 2.71 mm. CSA-Net demonstrated superior segmentation performance in the peripheral zone, urethra, and anterior fibromuscular stroma, and achieved second-best performance in segmenting the transition zone. The most significant improvement by CSA-Net was observed in the urethra, where it improved the Dice coefficient from 0.615 to 0.653 and reduced the Hausdorff distance from 2.08 mm to 1.43 mm, highlighting the model's proficiency in segmenting complex structures. Additionally, a more noticeable difference was observed between the performances of 2D and 2.5D methods in this challenging multi-class segmentation task (involving four different classes, with two being very small regions in the image). This highlights the benefits of 2.5D methods for challenging segmentation tasks. Improved multi-class segmentation of the prostate facilitates the development of methods for more precise staging and localization of prostate cancer on MRI.

\subsection{Qualitative Evaluation}
\begin{figure*}[!hbt]
\centerline{\includegraphics[width=\linewidth]{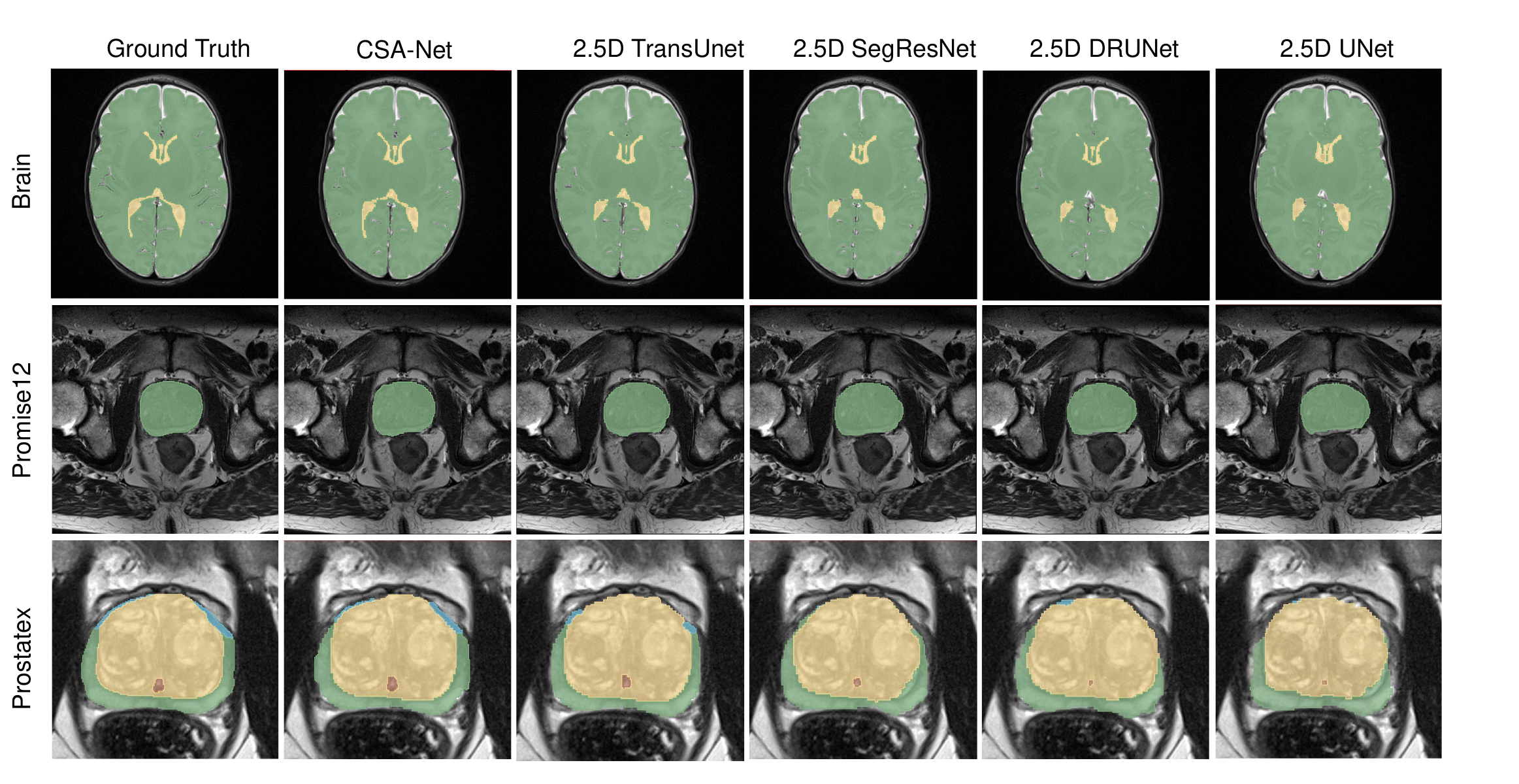}}
\vspace{-0.5em}
\caption{Segmentation results of 2.5D models on a representative subject from each of the three datasets. First row: green is the brain volume and yellow is the ventricles. Second row: green is the prostate capsule. Third row: yellow is the transition zone, green is the peripheral zone, orange is the urethra, and blue is the anterior fibromuscular stroma. Note, for the ProstateX dataset, the prostate is relatively small in the MRI; thus, we only showed a central region for better visualization of the  results.
}
\label{visualization}
\end{figure*}

Figure~\ref{visualization} illustrates the segmentation results of 2.5D models on three representative subjects, with each row representing a subject from one of the three datasets. For the Brain dataset, although all segmentation methods effectively segmented the brain volume, all methods except CSA-Net under-segmented the ventricles. For the Promise12 dataset, all segmentation methods successfully and accurately segmented the prostate capsule, except for the 2.5D UNet, which under-segmented the prostate. The results of CSA-Net are visually closest to the ground truth. For the ProstateX dataset, both 2.5D SegResNet and 2.5D UNet failed to segment the anterior fibromuscular stroma, while 2.5D TransUNet and 2.5D DRUNet significantly under-segmented this region. Additionally, all CNN-based methods under-segmented the urethra. The results highlight CSA-Net's capacility in addressing both binary and multi-class segmentation challenges, from uniformly shaped organs like the prostate to more complex anatomical structures such as the ventricles and the anterior fibromuscular stroma.

\section{Ablation Study}

\subsection{Importance of Key Model Components}

We investigated the importance of the following three key components in CSA-Net: the cross-slice attention module, the in-slice attention module, and the multi-head attention design. We reran CSA-Net using one of the following setups: (1) removing the cross-slice attention module; (2) removing the in-slice attention module; (3) replacing the multi-head attention design with a single-head attention design in both the cross-slice and in-slice attention modules.

\begin{table}[!hbt]
  \centering
  \caption{Impact of Key Components in CSA-Net. }
  \label{tb:ablation_key_components}
  \begin{tabular}{p{3cm} *{3}{c}} 
    \toprule
    \thead{Method} & \thead{Brain   \\DSC $\uparrow$} & \thead{ProstateX  \\DSC $\uparrow$}  & \thead{Promise12  \\DSC $\uparrow$}   \\
    \midrule
    w/o cross-slice attention & 0.886 & 0.651 & 0.911\\    
    w/o in-slice attention & 0.892 & 0.655 & 0.919 \\   
    single-head attention & 0.887 & 0.651 & 0.912\\
     CSA-Net & \textbf{0.896} & \textbf{0.659} & \textbf{0.921}\\
    \bottomrule
  \end{tabular}
\end{table}

Results in Table~\ref{tb:ablation_key_components} show that removing any key component from CSA-Net decreases segmentation accuracy, evidenced by reduced Dice coefficients across all three datasets. The most notable decline is observed with the removal of the cross-slice attention module, whereas excluding the in-slice attention module marginally impacts performance. Substituting multi-head with single-head attention leads to a significant drop in model performance, emphasizing the critical role of multi-head attention in analyzing information from multiple perspectives. Overall, this experiment underscores the important role of each CSA-Net component in achieving superior segmentation accuracy, where cross-slice attention improves contextual comprehension across slices, in-slice attention enables a more holistic understanding within a slice, and multi-head attention facilitates the capture of diverse features and relationships.

\subsection{Choice of the Number of Attention Heads}
The number of attention heads directly affects the CSA-Net model's ability to identify correlations between features in input image slices. More attention heads enable the detection of a broader range of data dependencies. However, additional attention heads increase the number of parameters, computational demands, and the risk of overfitting, particularly with inadequate regularization or limited data. We evaluated the performance of CSA-Net with various numbers of attention heads: 1, 4, 8, 16, and 32.

\begin{table}[!hbt]
  \centering
  \caption{Impact of the Number of Attention Heads. }
  \label{your_table_label}
  \begin{tabular}{>{\centering\arraybackslash}p{3cm} *{3}{c}} 
    \toprule
    \thead{Number of Heads} & \thead{Brain   \\DSC $\uparrow$} & \thead{ProstateX  \\DSC $\uparrow$}  & \thead{Promise12  \\DSC $\uparrow$}   \\
    \midrule
     1 & 0.887 & 0.651 & 0.912 \\
     4 & 0.888 & 0.652 & 0.913 \\
     8 & 0.892 & 0.654 &0.915\\
     16  & \textbf{0.896} & \textbf{0.659} & 0.921 \\
     32 & \textbf{0.896} & \textbf{0.659} & \textbf{0.922}\\
    \bottomrule
  \end{tabular}
\end{table}

Results in Table~\ref{your_table_label} show that increasing the number of attention heads in CSA-Net enhances segmentation performance across different datasets up to a certain point, with 16 heads emerging as the optimal configuration for balancing model complexity and accuracy. Beyond 16 heads, the marginal performance gains observed suggest a ceiling effect, indicating diminishing returns on further increases in attention head count. This finding underscores the importance of selecting an appropriate number of attention heads to maximize computational efficiency and minimize the risk of overfitting. Therefore, we chose 16 heads in this study as an optimal trade-off between performance and model simplicity.

\subsection{Center Slice as Query or Key-Value}
In the cross-slice attention module, choosing the center slice as either the query or the key-value pair affects information flow and segmentation results. When the center slice is used as the key-value pair, the attention mechanism prioritizes integrating contextual information from the neighboring slices into the representation of the center slice. This approach aims to enrich the center slice's features with additional context before segmentation. However, if the neighboring slices introduce noise or irrelevant details, it might adversely affect performance. Conversely, using the center slice as the query puts the emphasis on finding and emphasizing features in the neighboring slices that are most relevant to the center slice. This method can lead to more precise segmentation when the relevance of information varies significantly across slices.

\begin{table}[!hbt]
  \centering
  \caption{Impact of Using the Center Slice as a Query or Key-Value Pair.}
  \label{tb:center_slice_as_key_or_query}
  \begin{tabular}{p{3cm} *{3}{c}} 
    \toprule
    \thead{Method} & \thead{Brain  \\DSC $\uparrow$} & \thead{ProstateX  \\DSC $\uparrow$}  & \thead{Promise12  \\DSC $\uparrow$}   \\
    \midrule
    center slice as key-value & \textbf{0.896}  & \textbf{0.659} & \textbf{0.921}\\
    center slice as query & 0.892  & 0.636 & 0.914\\
    \bottomrule
  \end{tabular}
\end{table}

Results in Table~\ref{tb:center_slice_as_key_or_query} indicate that using the center slice as the key-value pair in the cross-slice attention module slightly outperforms its use as the query in all three segmentation tasks. This suggests that integrating contextual information from adjacent slices into the center slice's representation enhances segmentation accuracy. Benefiting from the spatial coherence in volumetric data, this configuration effectively enriches the center slice with valuable contextual insights.

\section{Discussion}

\subsection{2D vs 2.5D vs 3D Image Segmentation}

The selection of 2D, 2.5D, or 3D segmentation techniques depends on the specific segmentation task at hand. 2D image segmentation, characterized by its simplicity and low computational cost, operates on individual image slices. This method is particularly effective for tasks where inter-slice variability is minimal or when rapid processing is necessary. However, its major limitation is the lack of a comprehensive understanding of the spatial context, which may hinder its ability to precisely delineate the 3D characteristics of anatomical structures.
In contrast, 3D image segmentation exploits the spatial context within medical images by analyzing volumetric data. This method has the capability to capture the complex 3D architecture of tissues and organs, thereby enhancing segmentation accuracy, particularly for complex and closely interconnected anatomical structures. Nevertheless, the drawbacks of 3D segmentation include its higher computational requirements, limited availability of annotated data, and the complexity of data preprocessing procedures.
2.5D image segmentation provides a balance between 2D and 3D methods, especially beneficial for segmenting 2.5D images that exhibit low resolution in the through-plane direction. This method surpasses 2D techniques by integrating inter-slice spatial information, thereby enhancing the model's comprehension of 3D structures without incurring the full computational expense associated with 3D segmentation techniques.
Our experiments and results highlight the advantages of 2.5D methods for addressing segmentation challenges in 2.5D images.

\subsection{Clinical Implications}
\subsubsection{\textcolor{black}Segmentation of the Brain}
Accurate segmentation of the brain on MRIs leads to accurate estimations of brain volume. Volumetric data from term-equivalent MRIs allow bedside clinicians in the Neonatal Intensive Care Unit to study the impact of various treatments, such as nutrition, on brain development~\cite{valdes2024impact}. This information helps clinicians understand which treatment variables impact brain volumes, serving as a surrogate for neurodevelopmental outcomes~\cite{romberg2022mri}. There is a critical need to develop reliable, accurate, and rapid methods for obtaining whole-brain MRI volumes in this patient group, as manual segmentation is laborious and potentially less accurate~\cite{valdes2024impact}. CSA-Net offers clinicians a valuable tool for quickly analyzing entire brain volumes from MRIs in term-corrected premature neonates.

\subsubsection{\textcolor{black}Segmentation of the Prostate Capsule}
Accurate segmentation of the prostate capsule on MRI by CSA-Net significantly improves the precision and effectiveness of MRI-TRUS (transrectal ultrasound) guided biopsy procedures. The integration of MRI and TRUS imaging techniques heavily relies on accurate prostate segmentations within both modalities, and this improved boundary delineation ensures precise targeting of suspicious lesions identified on MRI, thus improving the diagnosis of clinically significant prostate cancer via biopsy. Furthermore, accurate prostate segmentation on MRI is essential for the development of machine learning models aimed at early detection of prostate cancer. This precise delineation allows machine learning models to concentrate on relevant information within the prostate during training, enabling them to identify subtle cancer imaging biomarkers that differentiate between benign and aggressive cancer tissues~\cite{seetharaman2021automated,bhattacharya2022selective}.

\subsubsection{\textcolor{black}Segmentation of Different Prostate Zones}
Accurate segmentation of the prostate into different zones allows radiologists and medical professionals to identify and localize lesions or abnormalities more accurately. Most prostate cancers begin in the peripheral zone. Segmenting the peripheral zone enhances tumor detection and localization due to the tumors' distinct appearance against the normal tissue in the peripheral zone on MRI. The transition zone often contains benign prostatic hyperplasia, which can resemble cancer. Therefore, segmenting the transition zone is crucial for differentiating between benign prostatic hyperplasia and cancer. Accurate segmentation of the urethra allows us to know its position relative to prostate lesions during procedures like radical prostatectomy, reducing the risk of urinary incontinence and complications. Although cancers in the anterior fibromuscular stroma are less common, they can be aggressive when present. Segmenting this area ensures a comprehensive prostate evaluation, preventing the oversight of potential lesions.

\subsection{\textcolor{black}Limitations and Future Directions}
Despite the promising results achieved by CSA-Net in addressing the unique challenges of 2.5D image segmentation, this study is not without its limitations. One of the primary constraints is the model's dependence on the quality of the neighboring slices, which can vary significantly in clinical datasets. This variability might affect the model's ability to consistently leverage inter-slice correlations, especially in datasets with resolution inconsistencies or artifacts. Furthermore, while CSA-Net exhibits lower computational demand compared to full 3D models, its balance of performance and efficiency has yet to be tested across a broader range of hardware configurations and in real-time clinical settings. Future directions for this work, including integrating CSA-Net with transfer learning or self-supervised learning approaches, could address the commonly faced challenge of training with limited labeled data. Investigating the application of CSA-Net to other medical imaging modalities, such as CT scans or 2.5D images in non-MRI contexts, could further validate its versatility and effectiveness.

\section{Conclusion}
In this paper, we introduced CSA-Net, a novel method for addressing 2.5D image segmentation challenges. CSA-Net incorporates a cross-slice attention module to capture 3D image features by learning correlations between image slices, and an in-slice attention module to capture global 2D image features by understanding the correlations between different regions within the center slice. Through extensive evaluations on three 2.5D segmentation datasets, we demonstrated that CSA-Net consistently outperforms leading 2D and 2.5D models.

\bibliography{ref.bib}
\bibliographystyle{IEEEtran}

\end{document}